\documentclass[12pt,preprint]{aastex}
\usepackage{mathrsfs}

\newcommand\beq{\begin{equation}}
\newcommand\eeq{\end{equation}}
\newcommand\bea{\begin{eqnarray}}
\newcommand\eea{\end{eqnarray}}
\newcommand\bal{\begin{align}}
\newcommand\eal{\end{align}}

\renewcommand\bal{\mbox{\boldmath$\alpha$}}

\newcommand{\wmap} {{\texttt{WMAP}}}
\newcommand{\one}{{\texttt{WMAP1}}}

\newcommand{\fiver} {{\texttt{WMAP5}}}

\newcommand{\kp} {{\texttt{ext}}}

\begin{document}

\title{Significant foreground unrelated non-acoustic
anisotropy on the one degree scale in WMAP 5-year observations}

\author{Bi-Zhu Jiang\altaffilmark{1,2}, Richard Lieu\altaffilmark{2},
Shuang-Nan Zhang\altaffilmark{1,2,3}, and Bart
Wakker\altaffilmark{4} }

\altaffiltext{1}{Physics Department and Center for Astrophysics,
Tsinghua University, Beijing 100084, China.}

\altaffiltext{2}{Department of Physics, University of Alabama,
Huntsville, AL 35899.}

\altaffiltext{3}{Key Laboratory of Particle Astrophysics, Institute
of High Energy Physics, Chinese Academy of Sciences, Beijing, China}

\altaffiltext{4}{Department of Astronomy, University of Wisconsin,
475 N. Charter St, Madison, WI 53706, USA}

\begin{abstract}

The spectral variation of the cosmic microwave background (CMB) as
observed by \wmap~was tested using foreground reduced \fiver~data,
by producing subtraction maps at the 1$^\circ$ angular resolution
between the two cosmological bands of V and W, for masked sky areas
that avoid the Galactic disk.  The resulting $V-W$ map revealed a
non-acoustic signal over and above the \fiver~pixel noise, with two
main properties. Firstly, it possesses quadrupole power at the
$\approx$ 1 $\mu K$ level which may be attributed to foreground
residuals. Second, it fluctuates also at all values of $\ell >$ 2,
especially on the $1^\circ$ scale ($200 \lesssim \ell \lesssim
300$). The behavior is {\it random and symmetrical} about zero
temperature with a r.m.s. amplitude of $\approx$ 7 $\mu K$, or 10 \%
of the maximum CMB anisotropy, which would require a `cosmic
conspiracy' among the foreground components if it is a consequence
of their existences.  Both anomalies must be properly diagnosed and
corrected if `precision cosmology' is the claim. The second anomaly
is, however, more interesting because it opens the question on
whether the CMB anisotropy genuinely represents primordial density
seeds.

\end{abstract}

\section{Introduction}

Studies of the cosmic microwave background (CMB, Penzias and Wilson
(1965)), the afterglow radiation of the Big Bang,  are currently in
a period of renaissance after the breakthrough discovery of
anisotropy by the \texttt{COBE} mission (Smoot et al (1992)).
Confirmed with much improved resolution and statistics by
\wmap~(Hinshaw et al (2009)), the phenomenon provides vital
information on the primordial `seeds' of structure formation. The
anisotropy is attributed to frequency shift of CMB light induced by
these `seed' density perturbations, which has the unique property
that it leads to changes in the temperature of the black body
spectrum and not the shape of it. The CMB has maximum anisotropy
power at the 1$^\circ$ scale, or harmonic number $\ell \approx$ 220,
with lower amplitude secondary and tertiary peaks at higher $\ell$.

The $\Lambda$CDM cosmological model (Spergel et al(2007)) explains
the entire power spectrum remarkably using six parameters, by
attributing the peaks to acoustic oscillations of baryon and dark
matter fluids, as long wavelength modes of density contrast enter
the horizon and undergo causal physical evolution.  CMB light
emitted from within an overdense region of the oscillation are
redshifted by a constant fractional amount, resulting in a cold
spot, which is a lowering by $\delta T$ of the black body
temperature $T$, and is frequency independent, i.e. $\delta T/T =
\delta\nu/\nu =$ constant. The opposite effect of blueshift applies
to underdense regions, leading to hot spots. Therefore, if the
anisotropy is genuinely due to acoustic oscillations, the inferred
change in $T$ at a given region should be the same for all the
`clean' frequency passbands of the \wmap~ mission. Since a
corresponding variation of the CMB flux $B(\nu, T)$ at any given
frequency $\nu$ is $\delta B = (\partial B/\partial T)\delta T$ if
the cause is solely $\delta T$ with no accompanying distortion of
the functional form of $B$ itself, the expected $\delta B$ at
constant $\delta T$ is then the `dipole spectrum' $\partial
B/\partial T$ which is well measured by \texttt{COBE-FIRAS} (Mather
et al(1994)). Moreover, the \wmap~data are calibrated w.r.t. this
dipole response.

A noteworthy point about the acoustic peaks is that one needs to
employ the technique of cross correlation to reduce the noise
contamination at high $\ell$, especially the harmonics of the second
and higher acoustic peaks.  Specifically one computes the all-sky
cross power spectrum \beq C_{\ell}^{ij} = \frac{1}{2\ell+1}
\sum_{m}~ a_{\ell m}^i a_{\ell m}^{j*}, \eeq where the indices $i$
and $j$ denote independent data streams with uncorrelated noise that
arise from a pair of maps at different frequency bands (or same band
but taken at different times), and $a_{\ell m}^i = \delta T_{\ell
m}^i$ is the apparent CMB temperature anisotropy for the spherical
harmonics $(\ell, m)$ as recorded by observation $i$.  Since the use
of multiple passbands is crucial to the accurate profiling of the
acoustic oscillations, it is important that we do compare them with
care, down to the level of measurement uncertainties.  Only {\it a
priori} statistically consistent maps should be cross correlated, in
the sense that any real discrepancies between such maps may carry
vital information about new physical processes that their cross
power spectrum does not reveal.  In one previous attempt to address
this point (see Figure 9 of Bennett et al(2003a)) \texttt{WMAP1}
data downgraded to an angular resolution commensurate with
\texttt{COBE}~were used to produce a difference (subtraction) map
between the two missions.  When displayed side by side with the map
of the expected noise for each resulting pixel, the two maps did
appear consistent. Nevertheless, this powerful method of probing the
CMB anisotropy does, in the context of the specific datasets used by
Bennett et al(2003a), suffer from one setback: it is limited by the
sensitivity and resolution of \texttt{COBE}.

In another test of a similar kind, we observe that each amplitude
$a_{\ell m}^i$ can further be factorized as $a_{\ell m}^i = a_{\ell
m} b_{\ell}^i$, where the array $b_{\ell}^i$ accounts for the
smoothing effects of both the beam and the finite sky map pixel
size, and $a_{\ell m} = \delta T_{\ell m}$ is the true amplitude of
the CMB anisotropy.  The results (see Figure 13 of Hinshaw et al
(2007)) indicate agreement of the variance $C_{\ell}^{ij}$, hence
$\delta T_{\ell}$, within the margin of a few percent for $\ell
\lesssim$ 400 among the many cross power spectra formed by the
various possible combinations of pairs of all-sky maps.  This offers
more ground for optimism, but to be definitive the remaining
discrepancy needs to be demonstrably attributed to noise,
instrumental systematics, or foreground emission.

The purpose of our investigation is to perform further, more
revealing comparisons than the two past ones described above,
initially by focussing upon the angular scale of the first acoustic
peak, which is $\sim$ 1$^\circ$. Our analysis will be done in both
real (angular) and harmonic domains, because while most of the
effort have hitherto been pursued in the latter, the former is the
domain in which the raw data were acquired and organized.

\section{The all sky difference map between the \fiver~V and W
bands}

We adopted the \texttt{Healpix}\footnote{See
\texttt{http://healpix.jpl.nasa.gov}.} pixelization scheme to ensure
that all pixels across the sky have the same area (or solid angle).
Firstly the W band data is smoothed to the V band resolution. Then
the whole sky map is downgraded to $\approx$ 1$^\circ$ diameter
(corresponding to \texttt{nside} of 64 in the parametrization of the
\wmap~database), which is not only commensurate with the scale of
global maximum $\delta T$ power, but also large enough to prevent
data over-sampling due to the use of too high a resolution, as the
size is comfortably bigger than the beam width of the \wmap~~V band
(61 GHz) larger than that of the W band.

\begin{figure}[H]
\centering
\includegraphics[scale=0.5,angle=90]{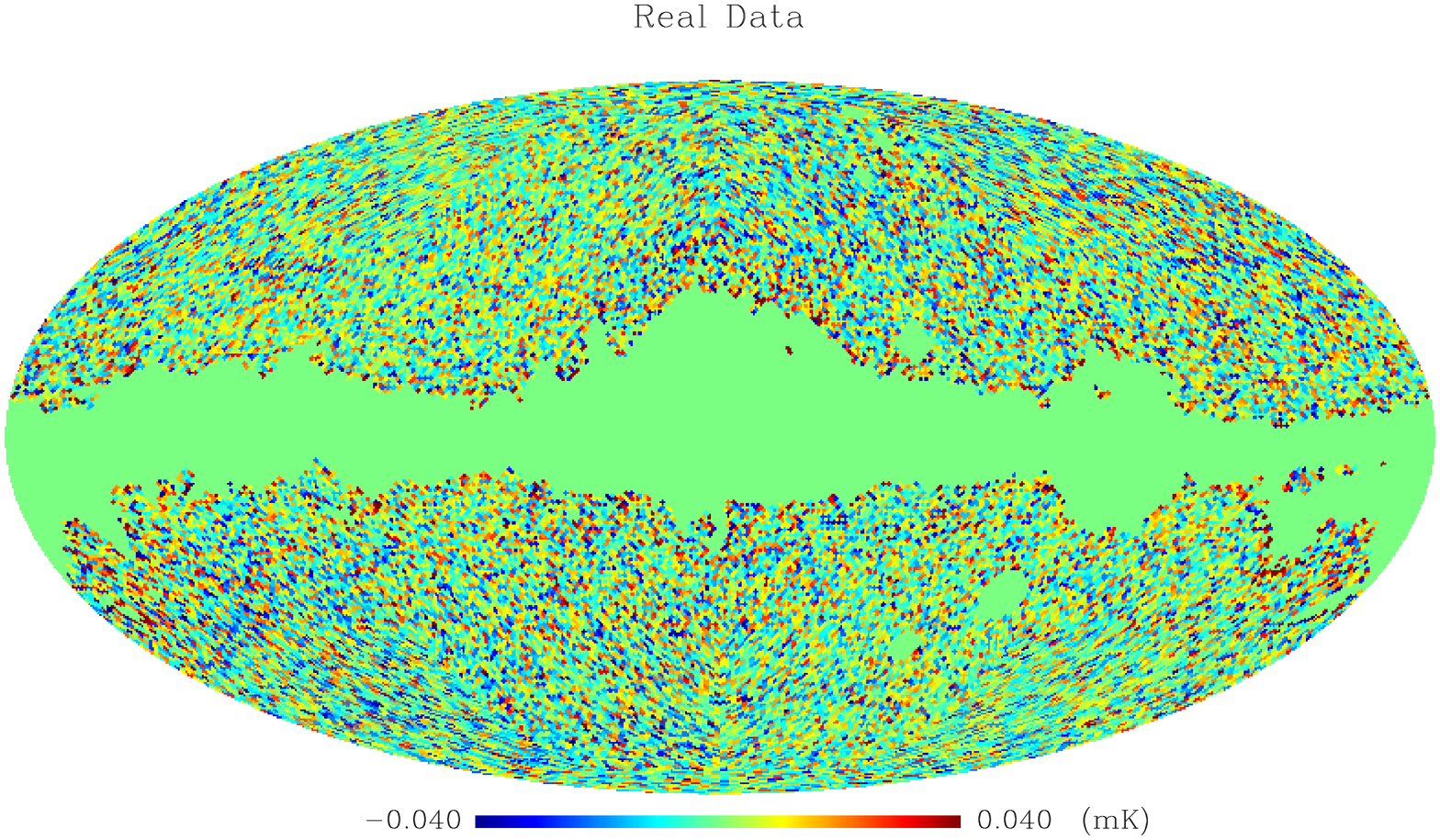}
\includegraphics[scale=0.5,angle=90]{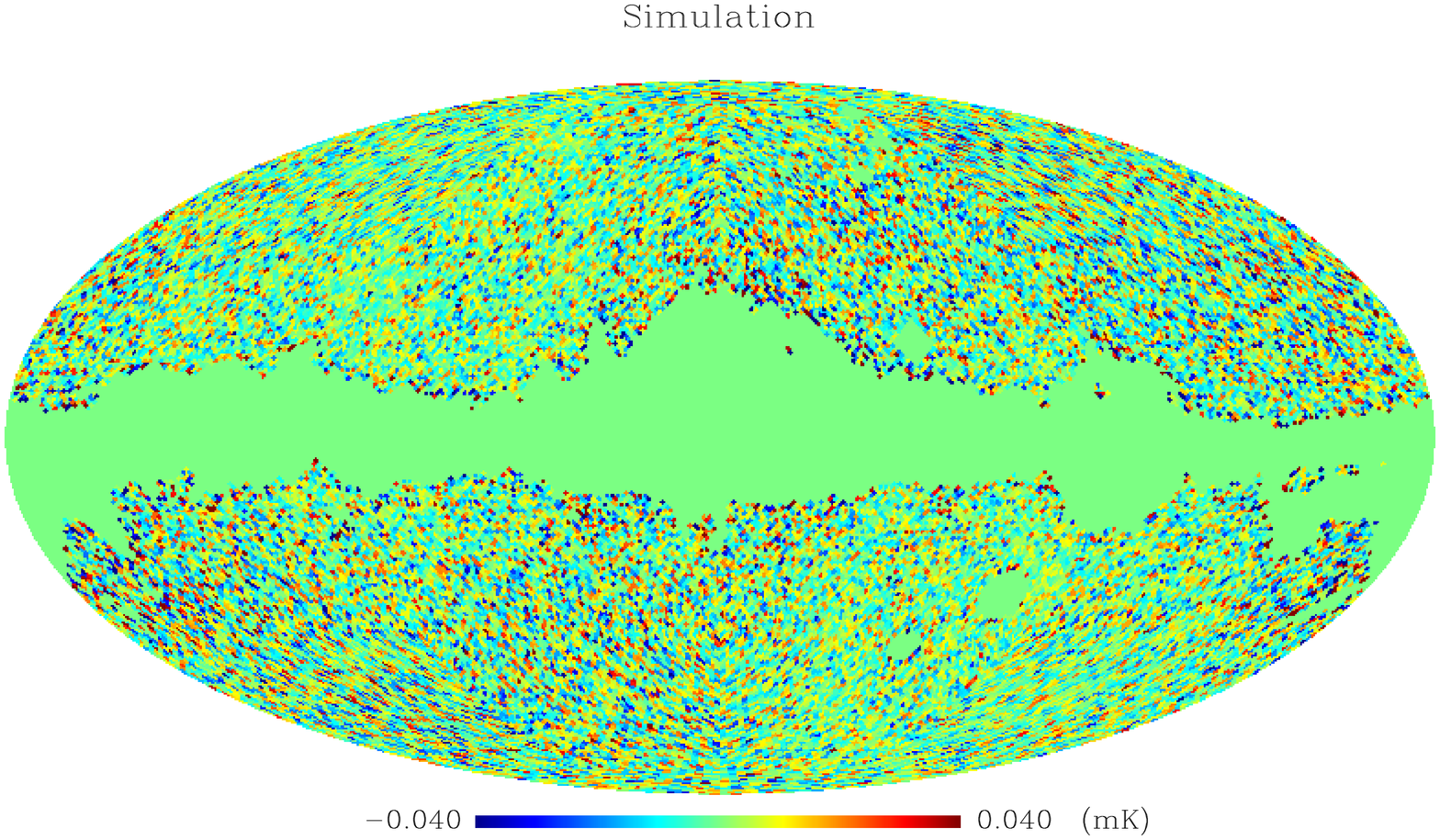}
\caption{The \kp-masked and point sources subtracted
\texttt{WMAP5}~$V-W$ map, viz. the difference map between the CMB
anisotropy as measured in the V band and the W band, for the real
data after the removal of residual monopole and dipole components
(top), and simulated pixel noise that reflect precisely the
observational condition (bottom). Both maps are plotted in Galactic
coordinates with the Galactic center $(l,b)=(0,0)$ in the middle and
Galactic longitude $l$ increasing to the left.  To avoid the
problems of beam size variation from one band to the next, the W
band data is smoothed to the V band resolution, then the pixels were
downgraded to the common resolution of \texttt{nside}$=$ 64 using
the foreground-reduced \fiver~data (see section 2); this resolution
under-samples the data in both bands. The color scale is coded
within a symmetrical range: those pixels with values beyond $\pm
40~\mu$K are displayed in the same (extreme) color; most of such
pixels are around the masked regions. The existence of additional
non- black body signal in the real data can readily be seen from
this comparison, as the simulated map is noticeably quieter.
\label{mapVWmK}}
\end{figure}

The resulting $\delta T$ values for the two cosmological passbands
of V and W, span $\approx$ 35,000 clean (i.e.
\kp-masked\footnote{\kp~is short for \texttt{external temperature
analysis}.} and foreground
subtracted\footnote{For foreground subtracted \fiver~maps see\\
\indent\texttt{http://lambda.gsfc.nasa.gov/product/map/current/m\_products.cfm}.})
pixels, from which a $V-W$ difference map at this $\approx$
1$^\circ$ resolution was made.  After removing the monopole and
dipole residuals (the latter aligned with the original \texttt{COBE}
dipole), this map is displayed in Figure \ref{mapVWmK} along with
the corresponding pixel noise map for reference; the latter
represents the expected appearance of the $V - W$ map if the CMB
anisotropy is genuinely acoustic in nature, so that the map would
consist only of null pixels should the \fiver~instruments that
acquired them be completely noise free.  When comparing the real
data map of Figure \ref{mapVWmK}a with the simulated map of Figure
\ref{mapVWmK}b, the former appears visibly noisier on the resolution
scale $\approx$ 1$^\circ$; moreover, the Leo and Aquarius (i.e. the
first and third) sky quadrants contain more cold pixels than the
other half of the sky, indicative of the existence of a quadrupole
residual.

The extra signals revealed by the $V - W$ subtraction map are
elucidated further in respect of their aforementioned properties by
examining the statistical distribution of the pixel values across
the four sky quadrants.  As shown in Figure \ref{hist_four_VW}, the
distribution of the 1$^\circ$ anisotropy is considerably wider than
that expected from the \fiver~pixel noise for all the quadrants, by
$\approx$ 10 $\mu K$, which is $\sim$ 10 \% of the $\approx$ 75 $\mu
K$ power in the first acoustic peak, and is therefore very
significant.   A detailed confirmation by Gaussian curve fitting is
given in Table \ref{VWmusigma}.

The $V - W$ quadrupole is more subtle, and is evident in the
residual plots at the bottom of each graph in Figure
\ref{hist_four_VW}, from which a slight skewness of the data to the
right is apparent in quadrants 1 and 3 (the quadrants of the CMB
dipole), with 2 and 4 exhibiting the opposite behavior.  For this
reason, the effect does not manifest itself as shifts in the
Gaussian mean value $\mu$ of Table \ref{VWmusigma}.  Rather, the
high statistical significance of both the quadrupole and the
degree-scale signals, with the former having a magnitude of
$\approx$ 1 $\mu K$, are established by computing the cross power
spectra of the temperature difference maps, Figure \ref{psVW}.  This
was performed at the resolution of \texttt{nside}$=$ 64 using the
\texttt{PolSpice} software\footnote{Available from
\texttt{http://www.planck.fr/article141.html}.}.  From Figure
\ref{psVW} also, the presence of excess non-acoustic anisotropy at
all harmonics $\ell > 2$, including the cosmologically important
$\theta\approx 1^\circ$ angular scale, appears robust. At the
$1^\circ$ scale ($200 \lesssim \ell \lesssim 300$), the r.m.s. is
about 7 $\mu K$, or 10 \% of the maximum CMB anisotropy. Lastly, the
$V-W$ quadrupole may be displayed in isolation by arranging the data
of the subtracted map as a multipole expansion \beq \delta T
(\theta,\phi) = \sum_{\ell,m} a_{\ell m} Y_{\ell m} (\theta, \phi),
\eeq  and evaluating at $\ell=2$ the amplitude \beq \delta T_\ell
(\theta,\phi) = \sum_m a_{\ell m} Y_{\ell m} (\theta, \phi), \eeq
(note $\delta T_\ell (\theta,\phi)$ is always a real number if the
original data $\delta T (\theta,\phi)$ are real).  The ensuing whole
sky map is in Figure \ref{VW_alm}, and the coordinates of the axes
are in Table \ref{poleaxis}.

\section{Interpretation of results}

The \fiver~$V-W$ map reveals two principal anomalies to be
explained: (a) the quadrupole at $\ell =2$, with an amplitude of
$\approx 1 \mu K$, and (b) the higher harmonic signals, especially
the $\approx 8~\mu K$ anisotropy at $\ell \gtrsim$ 200 (Figure
\ref{psVW}). Similar findings are also made by others, like the
noticeable hemispherical power asymmetry in the \one~analysis of
Eriksen et al (2004) and confirmed in the \fiver~data by Hoftuft et
al (2009), or the large scale distribution investigated by Diego et
al (2009).  Also because both (a) and (b) are not small effects,
claims to precision cosmology are overstatements until they are
properly accounted for and the cosmological model accordingly
adjusted.

Concerning (a), unlike the dipole, there is no previous known CMB
quadrupole of sufficient amplitude to justify its dismissal as a
cross band calibration residual. In fact, our reported amplitude of
$1~\mu$K is about 7 \% of the 211 $\mu$K$^2$ \fiver~anisotropy in
the unsubtracted maps of the individual bands at $\ell=$ 2, which is
far larger than the calibration uncertainty of $\approx$ 0.5 \%
(Hinshaw et al (2009)) for each band.

It will probably be more rewarding to search for remaining
foreground contamination not yet removed by the standard data
filtering and correction procedures of the \fiver~team (Bennett et
al(2003b), Gold et al(2009)).  Thermal dust emission might have a
power law spectrum with an index too close to that of the
Rayleigh-Jeans tail in the V and W bands for an appreciable V - W
signal, although this is an interesting scenario worthy of further
study (Diego et al 2009).  We consider here another possibility,
viz. free-free emission from High Velocity Clouds (HVCs, Wakker et
al (2009) and references therein). The clouds are moving at
velocities sufficiently large for any H$\alpha$ emission from them
to be outside the range\footnote{Example of a HVC missed by
\texttt{WHAM} is Hill et al 2009, a cloud of unit emission measure.
A notable exception (counter example) would be the HVC K-complex
(Haffner et al 2001), with an emission measure of 1.1 units, that
happens to fall inside the velocity window of \texttt{WHAM}.} of the
\texttt{WHAM} survey, the database employed to estimate the
free-free contribution to the \texttt{WMAP} foreground.  HVC
parameters for the larger and brighter clouds can reach: $n_e
\approx$ 0.2 $cm^{-3}$ and column density $\approx$
3~$\times$~10$^{19}$~cm$^{-2}$ (Wakker et al 2008).  This
corresponds to an emission measure of two units, or
6~$\times$~10$^{18}$~cm$^{-5}$, or $\approx$ 0.6~$\mu K$ of V-W
temperature excess (Finkbeiner D.P. (2003)), on par with the 1~$\mu
K$ of our observed quadrupole.  Moreover, as can be seen from the
all-sky map of $N_{{\rm HI}}$ and an estimate of the V-W excess in
Figure \ref{HVCs} when they are compared with Figures \ref{VW_alm}
and \ref{psVW}, the strength and distribution of HVCs do appear to
be responsible for a non-negligible fraction of the observed anomaly
on very large scales.  Further work in this area is clearly
necessary, and will be pursued in a future, separate paper.

We now turn to (b), the effect that occurs on the much smaller and
cosmologically most significant angular scale of 1$^\circ$.
Calibration issues are again immediately excluded here, since the 8
$\mu K$ anomalous amplitude is on par with the pixel noise of
\fiver~for the scale in question (Table \ref{VWmusigma}). Moreover,
because the subtracted $V - W$ dipole and the (unsubtracted) $V - W$
quadrupole, the latter being (a), are both relatively feeble
phenomena, of amplitudes $\approx$ 0.2 and 1 $\mu K$ respectively as
compared to the 7 $\mu K$ amplitude of (b), the prospect of smaller
scale fluctuations having been enhanced by a larger scale one can be
ruled out here.  CMB spectral distortion during the recombination
era, or subsequently from the Sunyaev-Zel\'dovich (SZ) scattering,
or from other foreground re-processing that were not properly
compensated by the data cleaning procedure of \fiver, could all be
responsible for the observed anomaly.   Although the first two
interactions (Sunyaev and Chluba (2008), Birkinshaw and Gull (1983))
exert much smaller influences than 7~$\mu K$ (bearing in mind that
the degree of SZ needs to be averaged over the scale of the whole
sky), the foreground could potentially play a relevant role in a
similar way as it did at very low $\ell$.  Thus, in respect of
free-free emission by HVCs alone, until a full survey at high
angular resolution is performed one cannot be certain that the
emission measure from these clouds is too weak to account for our
(b) anomaly.  However, the action of the foreground is {\it
systematic} in that it does {\it not lead to random and symmetric
temperature excursions} (about zero) between two frequencies of V
and W.  More precisely, because the sources or sinks involved have a
characteristic spectrum that differs from black body in a specific
way, any widening in Figure \ref{hist_four_VW} of the data
distribution w.r.t. the expected simulated gaussian ought to be
highly asymmetric.  This obviously contradicts our findings, i.e. we
note from Figure \ref{hist_four_VW} that the widening of the data
histogram is highly symmetric.  As a result, the symptoms do not
point to the foreground as responsible cause.

\section{Conclusion}

\begin{figure}[H]
\centering
\includegraphics[scale=0.3,angle=90]{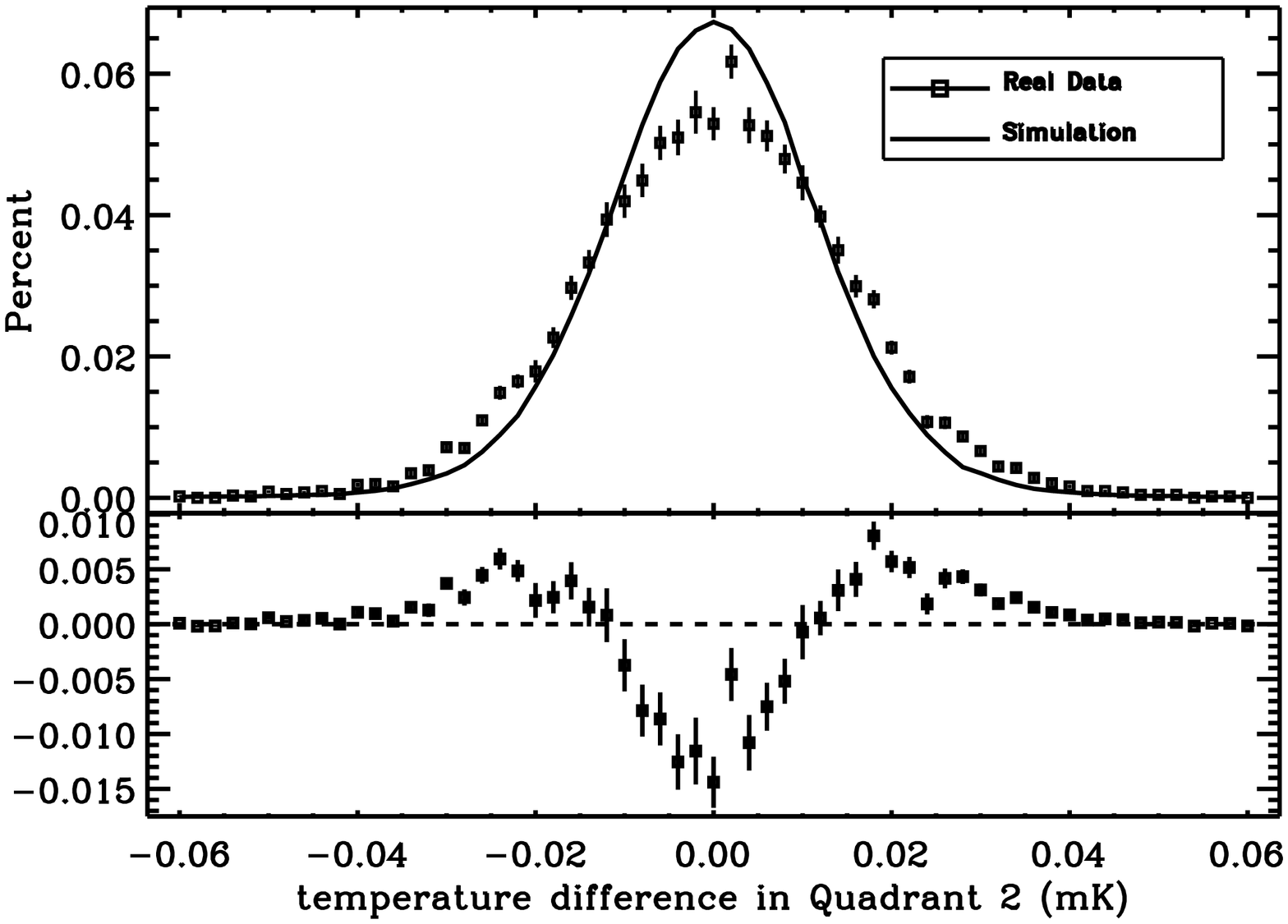}
\includegraphics[scale=0.3,angle=90]{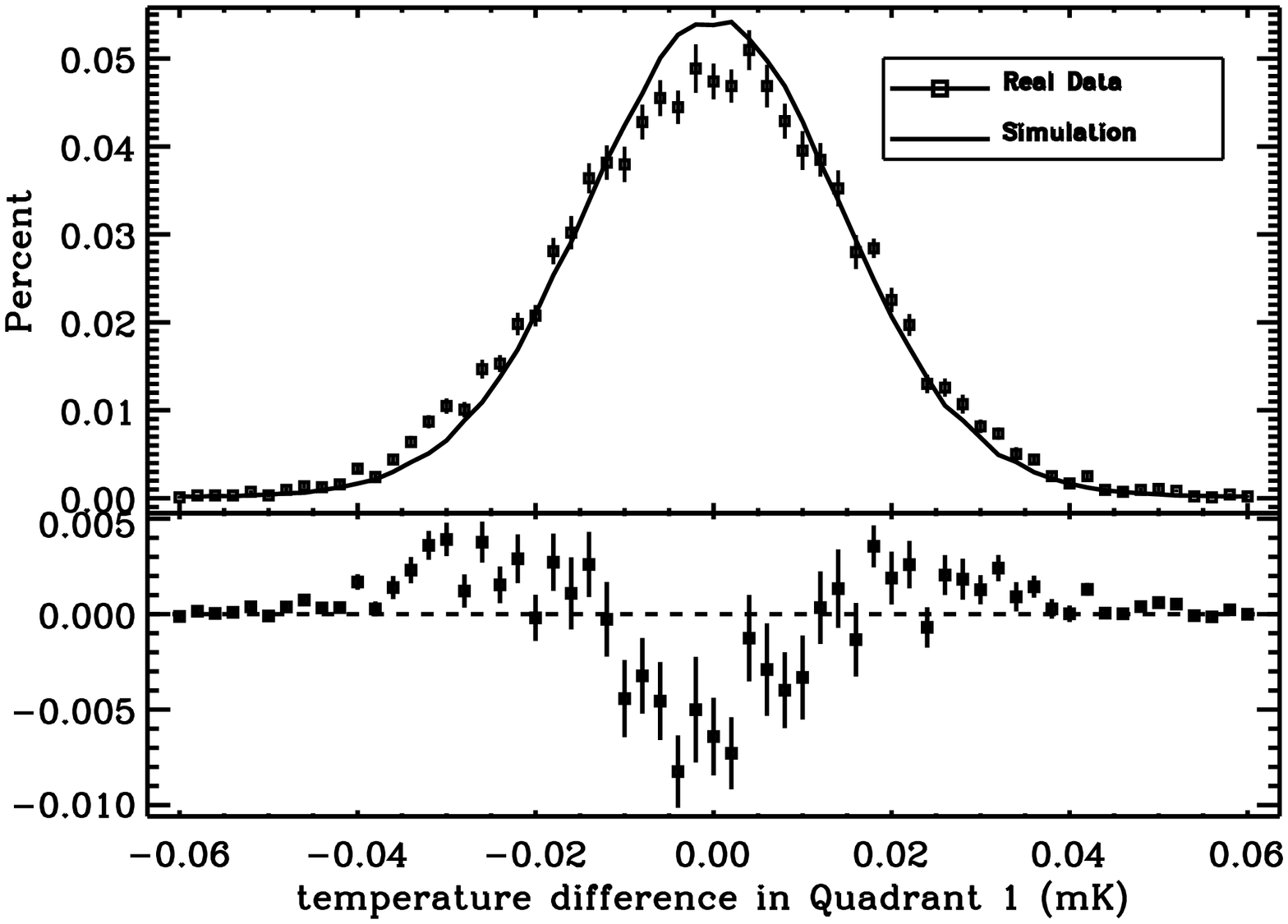}
\includegraphics[scale=0.3,angle=90]{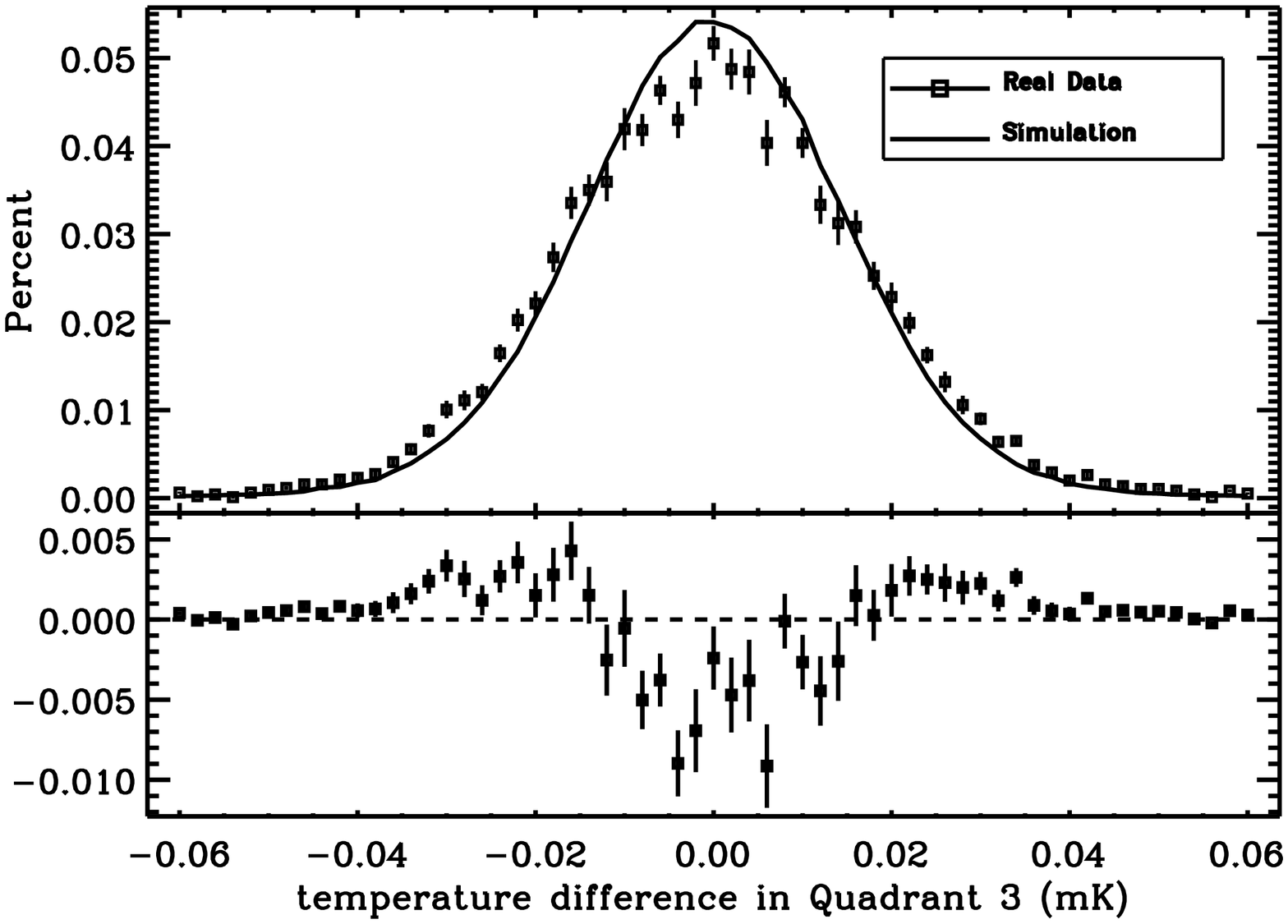}
\includegraphics[scale=0.3,angle=90]{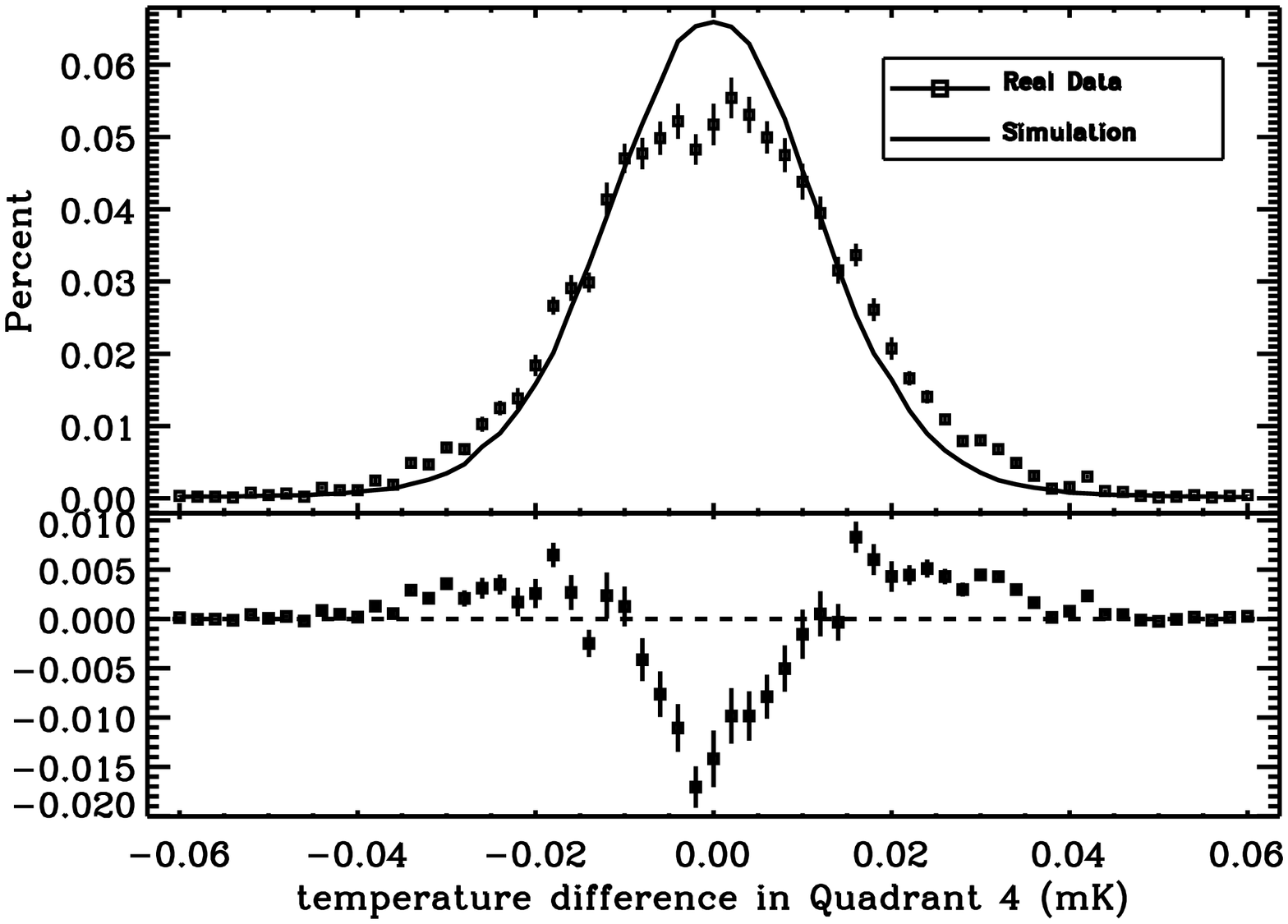}
\caption{The data points show quadrant sky occurrence frequency
distribution of the difference in the degree-scale
(\texttt{nside}$=64$) anisotropy between the \fiver~V and W bands,
while the errors in the data are due to the \fiver~pixel noise for
the same \kp-masked quadrant sky area, i.e. they are the statistical
fluctuations in the various parts of the solid line, which gives the
mean histogram of this noise.   The orientation of each quadrant
follows the same convention as the sky maps of Figure \ref{mapVWmK},
with the 1st and 3rd quadrants marking the \texttt{COBE} dipole.
\label{hist_four_VW}}
\end{figure}

\begin{figure}[H]
\centering
\includegraphics[scale=0.45,angle=90]{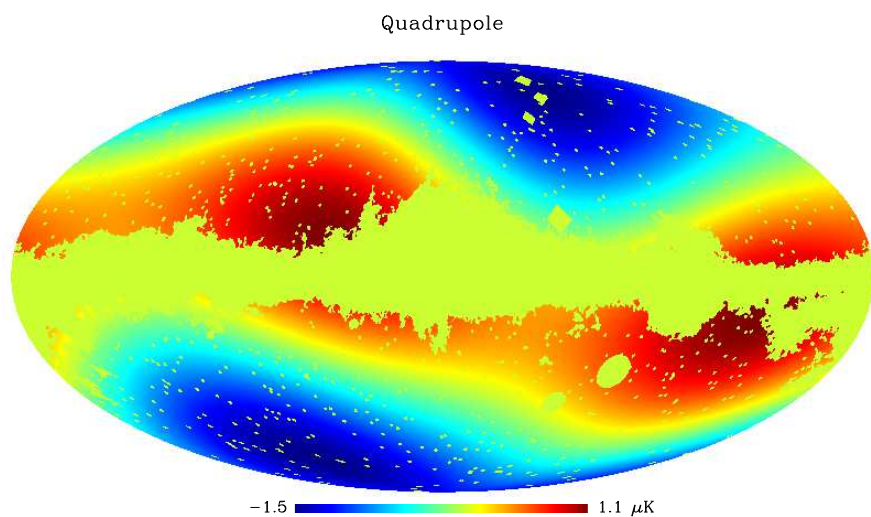}
\caption{$V-W$ quadrupole of the \texttt{nside}$=64$
\fiver~temperature difference maps, after \kp-masking and point
source subtraction. The mathematical procedure of extracting each
multipole $\ell$ is given in eqs. (3) and (4) of the text, and the
software used to do these computations was from \texttt{anafast} of
\texttt{Healpix}.\label{VW_alm}}
\end{figure}

\begin{figure}[H]
\centering
\includegraphics[scale=0.35]{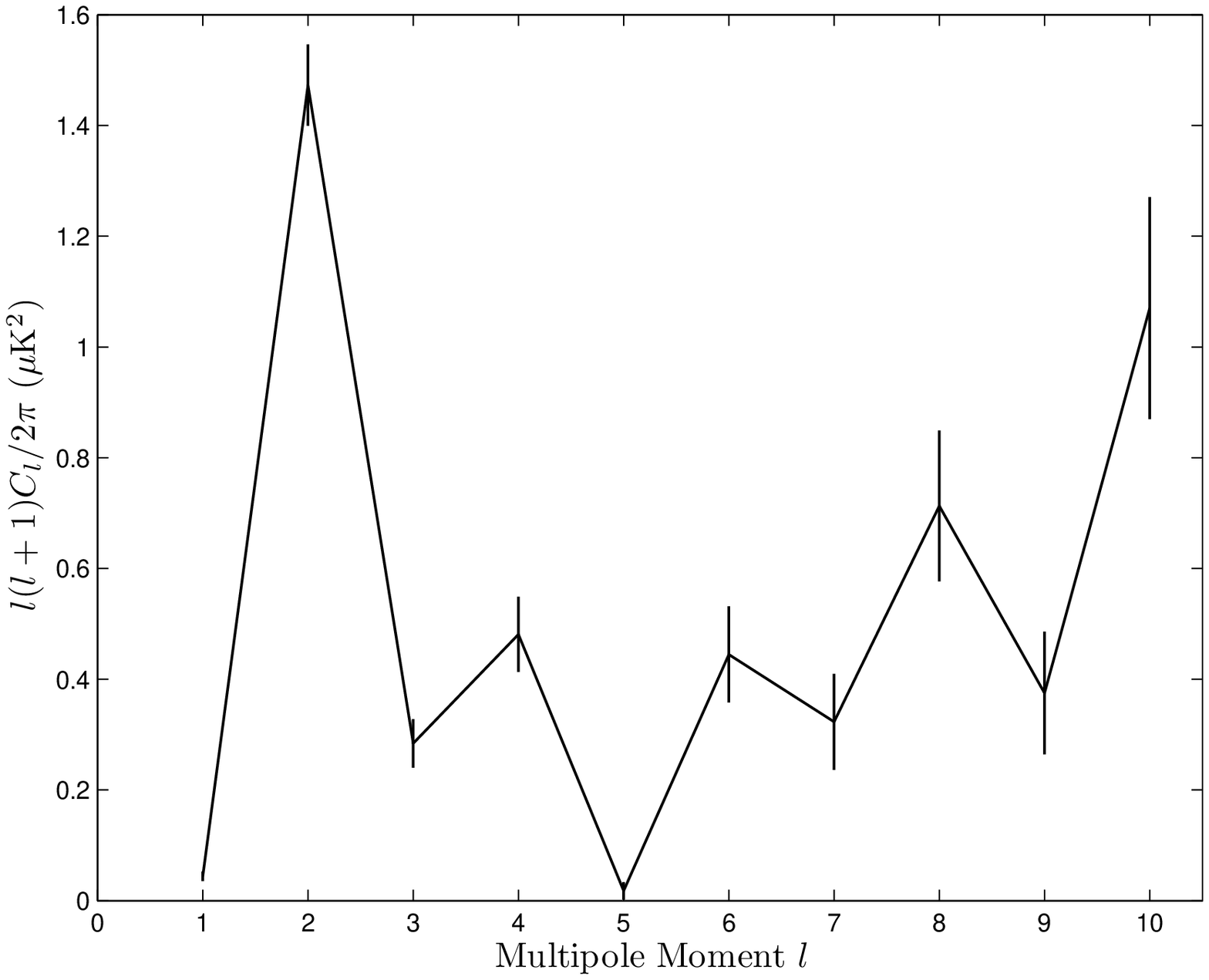}
\includegraphics[scale=0.35]{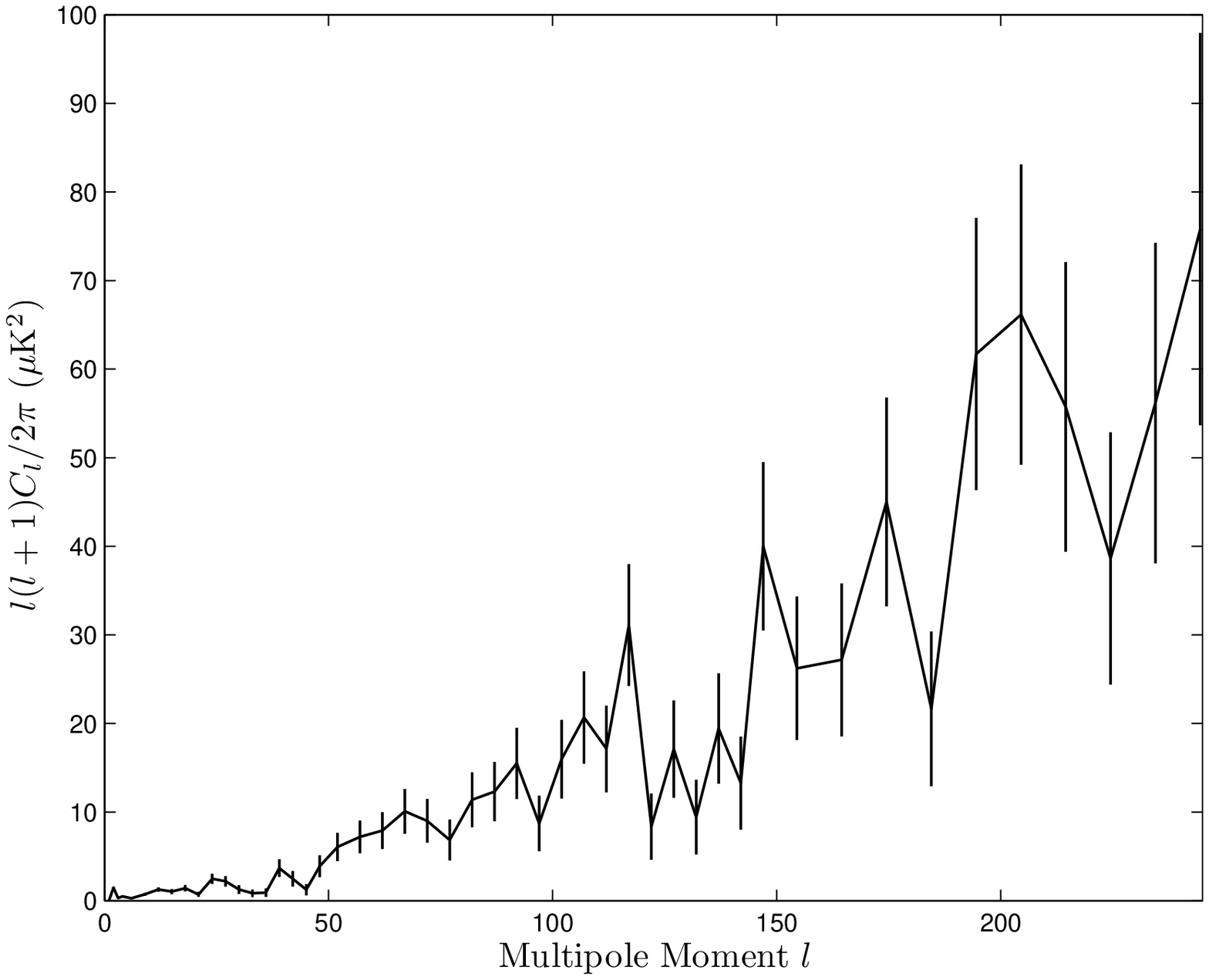}
\includegraphics[scale=0.35]{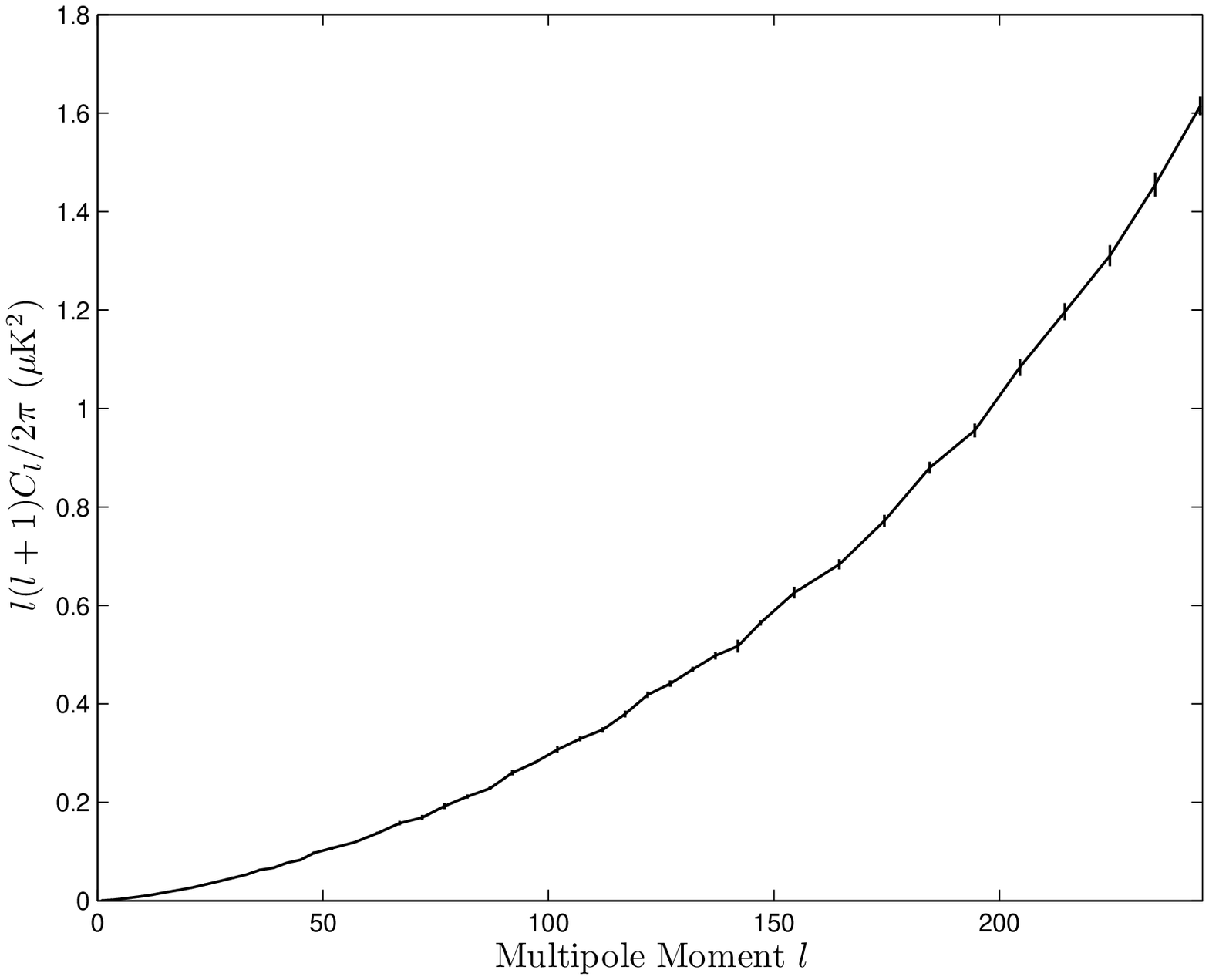}
\caption{Real and simulated (noise) power spectra of the
\fiver~$V-W$ map.   These are V-W {\it cross} power spectra computed
by cross correlating the first three years of observations with the
last two. The errors in the real data of the first two graphs
represent the pixel noise power of the last graph, i.e. 4c is the
average of 1,000 simulated realizations of the V-W \fiver~pixel
noise. Thus, if the noise power at harmonic $\ell$ is $(\delta
T_\ell)^2$ from 4c, the upper error bar in 4a and 4b will extend
from $T_l^2$ to $(T_\ell+\delta T_\ell)^2$ where $T_\ell$ is the
observed V-W anisotropy of each real data point (given by the
intersection of the error bars with the zig-zag line) in 4a and 4b.
The rising trend ($\sim l^2$) of all three curves towards higher $l$
simply reflects the relatively larger pixel noise for smaller
angular areas.  For $l>$ 200 the real data of 4a and 4b rapidly
become noise dominated. \label{psVW}}
\end{figure}

\begin{figure}[H]
\centering
\includegraphics[scale=0.6,angle=0]{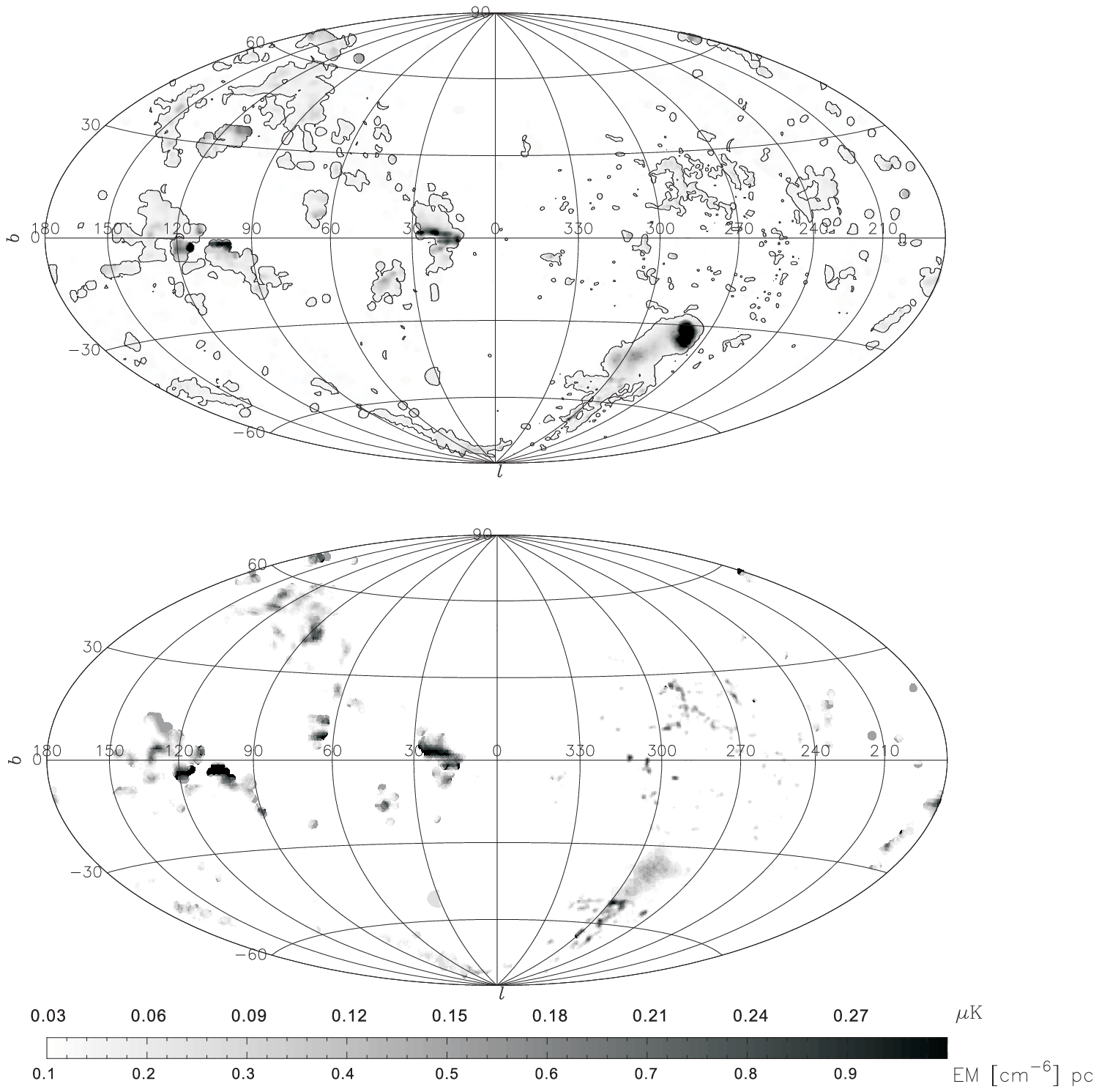}
\caption{Upper map shows 21 cm data of HVCs with HI column density
($N_{{\rm HI}}$) larger than 7 $\times$ 10$^{18}$ cm$^{-2}$ (i.e.
the greyscale shows $N_{{\rm HI}}$ with the outer contour at 7
$\times$ 10$^{18}$ cm$^{-2}$). Complex C is the cloud in the region
$l=$ 90$^\circ$ -- 130$^\circ$, $b=$ 40$^\circ$ -- 60$^\circ$.
Complex A is around $l=$ 150$^\circ$, $b=$ 30$^\circ$ -- 45$^\circ$.
The Magellanic Stream (MS) and Bridge is at $l=$ 280$^\circ$ --
310$^\circ$, $b <$ -30$^\circ$. The Leading Arm of the MS, plus some
other bright HVCs are at $l=$ 240$^\circ$ -- 300$^\circ$, $b=$
10$^\circ$ -- 30$^\circ$. Lower map gives our estimated V-W
temperature excess due to HVCs.  Note that because the dynamic range
of conversion from $N_{{\rm HI}}$ to this excess (via free-free
emission measure $EM$ of $N_{{\rm HII}}$) is not linear (e.g. Putman
et al 2003, Hill et al 2009).  Our approach is to assign 0.5 and 1.0
unit of $EM$, or 0.15 and 0.3 $\mu$K of V-W excess, to every
direction with $N_{{\rm HI}} \geq$ 2 $\times$ 10$^{19}$~cm$^{-2}$
and  5 $\times$ 10$^{19}$~cm$^{-2}$ respectively.   \label{HVCs}}
\end{figure}

We performed a new way of testing the black body nature of the CMB
degree scale anisotropy, by comparing the all-sky distribution of
temperature difference between the \fiver~cosmological bands of V
and W, with their expected pixel noise behavior taken fully into
consideration by means of simulated data.  In this way a non
acoustic signal is found in the \texttt{ext}-masked $V-W$ map at the
$\approx$ 1$^\circ$ resolution of \texttt{nside} $=$ 64, with the
following two properties. It has a quadrupole amplitude $\approx$ 1
$\mu$K (Figures \ref{hist_four_VW}, \ref{VW_alm}, and \ref{psVW})
which may in part be attributed to unsubstracted foreground
emission.  It also has excess anisotropy (or fluctuation) on all
scales $\ell
>$ 2, including and especially the scales of $200 \lesssim \ell
\lesssim 300$ where most of the acoustic power resides, and about
which the anomaly we reported is in the form of a symmetric random
excursion about zero temperature with a r.m.s. $\approx$ 8 $\mu K$
(Figures \ref{hist_four_VW} and \ref{psVW}, Table \ref{VWmusigma})
which is $\approx$ 10 \% of the maximum acoustic amplitude found at
$\ell \approx$ 220.   This type of excursion frustrates attempts to
explain the effect as foreground residuals, i.e. it opens the
question of whether the \texttt{WMAP} anisotropy on the 1$^\circ$
scale is genuinely related to the seeds of structure formation.

In any case, it is clear that both anomalies have sufficiently large
magnitudes to warrant their diagnoses through future, further
investigations, if the status of precision cosmology is to be
reinstated.

\begin{table}
\begin{center}
\begin{tabular}{|c|c|c|c|c|c|}
\hline
 \multicolumn{2}{|c|}{V - W} & $\mu (\mu$K)  & error ($\mu$K) & $\sigma$ ($\mu$K) & error ($\mu$K)\\
\hline
  & WMAP5  & -0.23 & 0.15 & 16.23 & 0.13 \\
\cline{2-6}
  Quadrant 1  & Simulation & 0.00 & 0.13 & 14.70 & 0.12\\
\cline{2-6}
  & Difference $\Delta$ & -0.23 & 0.20 & 6.88 & 0.40\\
\hline
  & WMAP5  & 0.24 & 0.12 &14.47 & 0.10\\
\cline{2-6}
  Quadrant 2  & Simulation & -0.04 & 0.12 & 12.10 & 0.10\\
\cline{2-6}
  & Difference $\Delta$ & 0.28 & 0.17 & 7.94 & 0.24\\
\hline
  & WMAP5  & -0.11 & 0.16 &16.22 & 0.13\\
\cline{2-6}
 Quadrant 3  & Simulation & 0.03 & 0.15 & 14.70 & 0.12\\
\cline{2-6}
  & Difference $\Delta$ & -0.14 & 0.22 & 6.86 & 0.40\\
\hline
  & WMAP5  & 0.40 & 0.13 &14.80 & 0.10\\
\cline{2-6}
 Quadrant 4  & Simulation & -0.01 & 0.13 & 12.26 & 0.10\\
\cline{2-6}
  & Difference $\Delta$ & 0.41 & 0.18 & 8.30 & 0.23 \\
\hline
\end{tabular}
\end{center}
\caption{Parameters for the gaussian curves that fitted the \fiver~
data and the pixel noise histograms (the latter are the solid lines)
of Figure \ref{hist_four_VW}.  Each parameter uncertainty is set by
the $\chi^2_{{\rm min}} + 1$ criterion, which represents the usual
68 \% (or unit standard deviation) confidence interval for one
interesting parameter, when the error bars shown in Figure
\ref{hist_four_VW} are employed for fitting both the real and pixel
noise data.  The difference in the width $\sigma$ between the two
models, which gives the distribution width of the additional random
signal, is given by $(\Delta\sigma)^2 = \sigma_r^2 - \sigma_s^2$.
The smaller simulated gaussian widths for quadrants 2 and 4
(relative to 1 and 3) is due to the higher exposure times there
(which contain the heavily scanned ecliptic poles) leading to lower
pixel noise.  \label{VWmusigma}}
\end{table}

\begin{table}
\begin{center}
\begin{tabular}{|c|c|c|}
\hline
   \multicolumn{3}{|c|} {V-W quadrupole location $(l,b)$} \\
\hline
hot & \multicolumn{2}{|c|} {$(-132.1^\circ,-14.4^\circ)$,$(48.0^\circ,14.4^\circ)$} \\
\hline
cold & \multicolumn{2}{|c|} {$(-81.5^\circ,68.0^\circ)$,$(98.5^\circ,-68.0^\circ)$} \\
\hline
\end{tabular}
\end{center}
\caption{Orientation of the quadrupole in the \fiver~ V-W map of
Figure \ref{VW_alm}. \label{poleaxis}}
\end{table}

\acknowledgments We are grateful to the referee for very valuable
suggestions towards the improvement of this paper.  Lyman Page,
Priscilla Frisch, Gary Zank, and Barry Welsh are also acknowledged
for helpful discussions. Some of the results were obtained by means
of the HEALPix package (G$\acute{o}$rski et al (2005)).

\end{document}